# Leakage Processes in Damaged Shale: *In Situ* Measurements of Permeability, CO$_2$-Sorption Behavior and Acoustic Properties

J. W. Carey, R. Pini, M. Prasad, L. P. Frash, and S. Kumar
Submitted as a chapter to the AGU monograph
"Caprock Integrity in Geological Storage"
in a section titled
"Geochemical Reactions in Fractured and Damaged Caprocks and Impacts to Permeability and Mechanical Stability"

## Abstract

Caprock integrity is one of the chief concerns in the successful development of a CO$_2$ storage site. In this chapter, we provide an overview of the permeability of fractured shale, the potential for mitigation of CO$_2$ leakage by sorption to shale, and the detection by acoustic methods of CO$_2$ infiltration into shale. Although significant concerns have been raised about the potential for induced seismicity to damage caprock, relatively little is known about the permeability of the damaged shale. We present a summary of recent experimental work that shows profound differences in permeability of up to three orders of magnitude between brittle and ductile fracture permeability. In the ductile regime, it is possible that shale caprock could accommodate deformation without a significant loss of CO$_2$ from the storage reservoir. In cases where CO$_2$ does migrate through damaged shale caprock, CO$_2$ sorption onto shale mineralogy may have a mitigating impact. Measured total storage capacities range from 1 to 45 kg-CO$_2$/tonne-shale. Once CO$_2$ is in the caprock, changes in the acoustic properties of CO$_2$-saturated shale that are predicted by Gassmann fluid substitution calculations show a significant reduction of the bulk modulus of CO$_2$-saturated shale.

## Introduction

During geologic sequestration of CO$_2$, most storage scenarios will require an impermeable geologic layer (commonly known as caprock) to prevent the escape of the buoyant supercritical CO$_2$ plume from the injection reservoir [Metz et al. 2005]. The most important types of caprock are shale and evaporite although low-permeability carbonates and sandstones can also serve as seals [Grunau 1987]. Methods of assessing caprock integrity have been developed as an important part of oil and gas exploration, where a good seal is as necessary to the occurrence of hydrocarbon deposits as the presence of good source rocks [Downey 1984; Grunau 1987]. The suitability and integrity of a caprock depends on a combination of low permeability, a capillary barrier to flow of gas and oil (i.e. the 'non-wetting' fluid phases), and the overall continuity and thickness of the formation [Downey 1984; Grunau 1987]. Ideally, the caprock should not have been subjected either to tectonic events or a production history that would have created faults and fractures [e.g., Hawkes et al. 2005].

$CO_2$ injection into reservoirs will change the state of stress in both the reservoir and caprock formations due to poro-mechanical, thermal and chemical effects. As a result of induced stress, the caprock can be mechanically damaged, pre-existing sealing faults and fractures can be re-activated, or new fracture systems can be created. Such changes are of particular concern when they impact the caprock because they could initiate fluid migration out of the storage reservoir.

For these reasons, recent work has called into the question the viability of $CO_2$ storage in the light of potential injection-induced damage. Zoback and Gorelick [2012] observe that in many places the earth's crust is "critically stressed" such that injection of $CO_2$ would result in small-magnitude earthquakes. The point at issue is that, despite their relatively low magnitude, these events could result in fault activation across the caprock and compromise storage security. Others have disagreed with this conclusion on the basis of earthquake locations, the historical record of oil and gas reservoirs, and considerations related to coupled mechanics and fluid flow [Juanes et al. 2012; Villarrasa and Carrera 2015]. Nevertheless, the threat of injection-induced earthquakes remains a significant concern for the risk assessment of a $CO_2$ sequestration operation.

This debate, however, has not addressed the potential *consequences* of the presence of fractures through caprock. In order for damage to the caprock to have an impact, a significant and persistent flow of $CO_2$ must occur. In this work, we focus on the question of permeability and transport of $CO_2$ through fractured shale. We consider a scenario in which injection processes, natural tectonic activity, or perhaps even an existing damage zone have created a potential pathway for $CO_2$ migration from the reservoir through a shale caprock. We consider how the coupled processes involved in geomechanical damage might affect leakage risk, namely enhanced permeability, (immiscible) $CO_2$ flow, fracture closure, and $CO_2$ trapping by sorption.

Likewise, there are few studies that have developed methods of detecting damaged caprock or the presence of $CO_2$-saturated caprock. $CO_2$ flow into and through caprock will necessarily be accompanied by fluid exchange. Such processes may be associated with a change in seismic and electrical properties. We consider the magnitude of changes in these geophysical properties that might be used to monitor caprock damage. Potential changes of geophysical properties can be attributed to at least two effects: the effect of a change of saturating fluid from connate water or hydrocarbon to $CO_2$ and the effect of geomechanical damage.

This chapter consists of a literature review of what is known of the factors governing potential leakage of $CO_2$ through shale caprock, focusing on the properties of fractured or damaged shale. We discuss three sub-topics in more detail based on recent experimental work by the authors: 1) fracture-permeability behavior of shale; 2) sorption of $CO_2$ in shale; and 3) acoustic properties of $CO_2$- reacted shale. Other recent reviews of caprock integrity and $CO_2$ sequestration include those of Fitts and Peters [2013]; Shukla et al. [2010]; Song and Zhang [2013] and a recent review of the geomechanical stability of caprock by IEAGHG [2015].

## Caprock Integrity Failure

A good caprock isolates water-immiscible, buoyant fluids (supercritical $CO_2$, oil and gas) in the subsurface. There are clearly many examples of good caprock as illustrated by the many accumulations of oil, gas and even $CO_2$ (e.g., $CO_2$ reservoirs of the southwestern US, Allis et al. 2001]. However, caprock might also have poor integrity and could have allowed fluids to escape as evidenced by subsea oil leaks [Hornafius et al. 1999], leaking oil reservoirs [Macgregor 1996], and travertine surficial deposits formed above $CO_2$ reservoirs [Keating et al. 2014; Moore et al. 2005]. Identifying good caprock involves establishing the basic geometric features of a caprock (vertical and lateral containment of the reservoir; Biddle and Wielchowsky 1994], determining that the capillary properties of the caprock will block entry of the immiscible phase of interest; and determining that stresses arising from tectonic and/or hydrocarbon production operations have not previously damaged the caprock [Cartwright et al. 2007; Hawkes et al. 2005; Macgregor 1996].

In studies of caprock integrity for $CO_2$ sequestration, much work has focused on measuring caprock permeability and on assessing the quality of the capillary barriers of undamaged shale [e.g., Armitage et al. 2011; Heath et al. 2012; Hildenbrand et al. 2002; Schlömer and Kroos 1997; Song and Zhang 2013; Wollenweber et al. 2010]. Typical absolute permeability values range from 0.001 to 1 µD [e.g., Hildenbrand et al. 2002]. However, as noted by Hildenbrand et al. [2002] and others, injection pressures must exceed the capillary breakthrough pressure before $CO_2$ flow can occur (from 0.1 to 6.7 MPa for the samples considered in Hildenbrand et al. 2002]. Values of the capillary breakthrough pressure can be used to compute the maximum thickness of a $CO_2$ plume beneath a seal such that breakthrough of the gas phase is not possible [e.g., Naylor et al. 2011]. Thus, limiting the thickness of the $CO_2$ plume limits the potential of transport processes of $CO_2$ through the caprock to only diffusion.

Much less work has been done on the permeability of *fractured* shale and, in particular, on the consequences of caprock fractures (i.e., by measuring the permeability of the resulting fracture system) either at the field scale [Cartwright et al. 2007] or laboratory scale [Carey et al. 2015; Gutierrez et al. 2000]. Some studies indicate that the predominant risk for caprock integrity is fracture flow [e.g., Edlmann et al. 2013; Pawar et al. 2015]. Interestingly, Skurtveit et al. [2012] find that even in studies of capillary breakthrough of $CO_2$ through nominally "intact" shale samples, deformation occurs as a result of dilation and microfracture formation.

**Origin of faults in reservoir-caprock systems**

A critical operational parameter for $CO_2$ storage is the maximum sustainable reservoir overpressure that can be achieved without fault activation [Sibson 2003]. Injection into a reservoir can reactivate faults due to an increase of pore pressure within a fault that reduces the effective stress normal to the fault plane. Pore pressure changes within a reservoir will also modify the total stress state as shown by the significant changes to the horizontal minimum stress reported for measurements in the field [Addis 1997]. These changes would tend to increase fault slip potential in some normal fault regimes and most thrust fault environments [Hawkes et al. 2005]. Faults in the caprock can also be induced by the dilation or compaction of the underlying reservoir in response to injection or production, which is likely to be particularly pronounced for elastically soft reservoirs [Hawkes et al. 2005]. While the reservoir experiences a change in pore-pressure that results in changes to the horizontal stress, the caprock does not; this can lead to the development of shear stresses at the interface between the reservoir and the caprock, which could also damage wells [Hawkes et al. 2005]. Hydraulic fractures are an additional mode of potential failure; these could originate due to pore-pressure at the top of the reservoir exceeding the least principle stress in the caprock [e.g., Finkbeiner et al. 2001]. In general, the formation of hydraulic fractures in the caprock would only be possible in the absence of pre-existing, low-cohesion fractures of suitable orientation [Morris et al. 1996; Sibson 2003]. Such fractures can be more easily stimulated at lower injection pressures than hydraulic fractures.

The link between fractures and fluid flow through caprock is well established [e.g., Hickman et al. 1995]. Several studies on hydrocarbon migration have found that fluid pressure in such reservoirs is bounded by the minimum horizontal stress; this suggests that an equilibrium is maintained in such hydrocarbon systems through episodic faulting and fluid migration [Dewhurst et al. 1999; Finkbeiner et al. 2001; Hermanrud and Bols 2002; Hunt 1990; Ingram and Urai 1999; Watts 1987]. A useful concept put forward by Finkbeiner et al. [2001] is that of a "dynamic capacity model" representing the critical pore pressure at which a caprock fails either by activation of an existing fault or by a newly formed hydraulic fracture into the caprock. As discussed above, this critical pore pressure would generally be lower in the case of fault activation. Given the arguments that the Earth's crust is critically stressed with abundant fractures [Zoback 2007], this further suggests that the limits of injection pressure would be lower than those set based on criteria for hydraulic fracture failure of the caprock. Such behavior is reported in Townsend and Zoback [2000] who provide evidence that much of the crust maintains a hydrostatic pore-pressure due in part to fault activation that allows redistribution of fluids [also see Sibson 2003].

The picture for fracture development in caprock is complex. On the one hand, it is clear that shale caprock provides long-term sealing behavior for oil and gas reservoirs and that therefore it could provide similar storage for $CO_2$. On the other hand, it is also clear that reservoirs leak and that leakage can be initiated by fluid pressure-induced fault activation. The two logical conclusions from this are that (1) injection pressures that can trigger faults have to be avoided a priori, and that (2) the potential leakage of $CO_2$ along faults (either because of existing faults or stimulated faults) has to be considered as part of the risk assessment of a storage operation. *The missing component in this analysis is a consideration of the consequences of leakage on fractures or faults*: that is, what is the permeability of faults and what do we know of the time-dependence of permeability and the effect of the particular properties of supercritical $CO_2$ on fault transmissivity? Such information is key to the development of contingency plans that have to be put in place in case such an event occurs.

**Permeability of Fractured and Faulted Shale**
Faults can be transmissive, sealing, vertically transmissive but laterally sealed, and episodic in behavior [Aydin 2000]. The occurrence of fault-compartmentalized oil reservoirs clearly shows that the mere existence of faults does not preclude long-term accumulation of buoyant hydrocarbon or supercritical $CO_2$ for that matter [e.g., Fisher and Knipe 2001]. In fault-bounded reservoirs, the most important sealing mechanism is the juxtaposition of an impermeable unit against the reservoir although the faults themselves may act as seals [Downey 1984; Fisher and Knipe 2001]. Several methods of predicting fault transmissivity are based on clay mineralogy and include clay smear potential, shale smear factor, and shale gouge ratio, as summarized in Edlmann et al. [2013] and Fisher and Knipe [2001], in which increasing clay content promotes sealing behavior.

Faults have complex structures that are often multi-stranded and may consist of a fault core surrounded by a damage zone. Adyin's [2000] review of fault structures indicates that fault cores in porous sandstone, consisting of comminuted material, are typically less permeable than the original rock by up to four orders of magnitude, whereas the damage zone might have enhanced permeability up to two orders of magnitude. For example, Agosta et al. [2007] shows that fault zones in carbonate host rocks can be separated into two main units with distinct hydraulic characteristics: a sealing fault core that prevents fluids from crossing the fault, and a cracked zone surrounding the core that allows fluid flow parallel to the fault. The latter can also occur due to the formation of opening-mode fractures within the cemented fault rocks. In Vialle et al.'s [2016] discussion of fault structures, they describe work in low-porosity rocks that shows contrasting behavior with an elevated permeability fault core and reduced permeability damage zone. Aydin [2014] reviews fault structures in shale including the likely low permeability of shear bands. The key message here is that the hydraulic properties of faults are likely to be highly anisotropic, with faults in some cases acting as barriers and in other cases conduits to flow.

The permeability resulting from shale deformation is likely to be strongly influenced by whether strain is accommodated by brittle versus ductile behavior [Dewhurst et al. 1999; Ingram and Urai 1999; Nygård et al. 2006]. Brittle shear deformation is characterized by dilation and the possibility of formation of fracture apertures, thus increasing permeability. On the contrary, ductile behavior is characterized by compaction (net porosity reduction) and would likely result in more limited changes or even reduction in permeability. This transition from dilation to contraction can be expressed by the concept of critical state behavior of deforming sediments [Schofield and Wroth 1968; Jones and Addis 1985]. Two approaches to predict critical state behavior are based on the concept of the overconsolidation ratio (OCR; defined by the ratio of the maximum vertical stress experienced by the shale to the present-day vertical stress) and the brittleness index (defined in terms of unconfined strength; Ingram and Urai 1999; Nygård et al. 2006]. (We note that there are many different definitions of brittleness index in the literature). Nygård et al. [2006] find the transition to ductile behavior at OCR > 2.5, suggesting that brittle behavior is observed in shale that is at substantially shallower depths in relation to the maximum depth of burial.

The mineralogy of shale must also play a role in ductility and fracture transmissivity. Bourg [2015] argues that there is a critical transition in brittle to ductile behavior in shale as clay content increases above 33%. In support of this concept, Bourg [2015] found that shale gas reservoirs had lower clay content than shale formations proposed for radioactive waste repositories and $CO_2$ sequestration. This is also consistent with experimental observations including those of Sone and Zoback [2013] who found increasing ductility and visco-elastic creep with clay content of shale gas samples. Creep is likely to play an essential role in understanding the long-term permeability of fractured shale with such self-sealing processes a focus of work in the European nuclear waste repository community [e.g., Bock et al. 2010].

Measurements of permeability of fracture-damaged shale are relatively scarce perhaps because of challenges posed by anisotropy, heterogeneity and sensitivity of shale to moisture . Some work has focused on investigations of manufactured fractures or samples with natural fractures [e.g., Bernier et al., 2007; Cho et al., 2013; Davy et al., 2007; Edlmann et al. 2013; Gutierrez et al., 2000; Zhang, 2013; Zhang et al., 2013]. For example, Edlmann et al. [2013] showed that a shale sample with natural fractures would not transmit supercritical $CO_2$, but was permeable to gaseous $CO_2$, which they attributed to increased fracture apertures at lower confining pressure.

An even smaller set of laboratory studies have made measurements of shale fractured at *in situ* conditions in order to gain insight into fracture-permeability behavior at reservoir conditions [Bernier et al. 2007; Carey et al. 2015; Menaceur et al. 2015; Monfared et al. 2012; Nygård et al. 2006; Zhang and Rothfuchs 2008]. Bernier et al. [2007] made measurements of hydraulic fracture conductivity using a hollow cylinder method and observed an increase of permeability of 4-5 orders of magnitude before self-sealing processes reduced these values. In triaxial compression experiments, Bernier et al. [2007] observed a *decrease* in permeability of Boom clay, but a significant increase (unspecified) in the permeability of Opalinus clay. In similar tests, Nygard et al. [2006] and Zhang and Rothfuchs [2008] also observed increases in permeability, which the former did not quantify and the latter observed a 4 order of magnitude increase to 1 µD. Menaceur et al. [2015] and Monfared et al. [2012] used a hollow-cylinder method on Callovo-Oxfordian and Boom shale, respectively, and found no increase in permeability after fracture formation. Results by Carey et al. [2015] using direct-shear methods will be discussed below.

Field measurements of fault permeability in shale were conducted by Nussbaum and Bossart [2008] for the Main Fault in the Mont Terri underground rock laboratory. They found that the fault zone had permeability similar to the impermeable Opalinus clay matrix. Guglielmi et al. [2015] reactivated the strike-slip Tournemire Fault by injection of fluid using a straddle packer system. They calculate a 2-order of magnitude increase in permeability following fault reactivation.

**Geochemical interaction of $CO_2$ with fractured shale**
The thickness and impermeability of shale indicate that $CO_2$-induced alteration of the bulk material is unlikely to lead to the formation of significant leakage pathways [e.g., Fitts and Peters 2013]. This is not to say that $CO_2$ has no effect on shale: Armitage et al. [2013] found that flow of $CO_2$-saturated water through a chlorite-siderite cemented siltstone resulted in an 8-fold increase in permeability from about $10^{-20}$ to $10^{-19}$ m². However, the most important geochemical effects on fracture permeability are likely to be the result of dissolution and precipitation processes [e.g., Detwiler 2008]. Under geological confinement, dissolution can lead to either fracture closure as asperities collapse or localized aperture opening (channel formation) depending on relative rates of chemical reaction and fluid movement [Detwiler 2008]. There has been relatively little work on the coupled geochemical-hydrologic effects of flow of $CO_2$ and brine on fractured caprock. Most of such experimental work has focused on fractures in carbonate [e.g., Ellis et al. 2011; Elkhoury et al. 2013; Noiriel et al. 2009].

In addition to mineral reactions, the geochemistry of $CO_2$ sorption may be an important component of $CO_2$ migration through shale. According to Bourg et al. [2015], the role of $CO_2$ sorption is poorly understood. Studies of $CO_2$ sorption on shale include those of Heller and Zoback [2014] who measured $CO_2$ sorption up to a pressure of 4 MPa and found that it was 2-3 times greater than methane (12-75 SCF $CH_4$/ton, i.e. 0.25-1.6 kg $CH_4$/tonne) on 4 different gas shale samples. This is much lower than the approximately 770 SCF $CO_2$/ton (44 kg/tonne) measured by Busch et al. [2008] at a pressure of 12 MPa on Maderung shale. We provide additional discussion of sorption processes and their role in potentially mitigating $CO_2$ leakage through caprock below.

Finally, $CO_2$ has the potential of altering fracture development and propagation through chemical reactions or stress corrosion [Anderson and Grew 1977; also see a recent review of the effect of $CO_2$ on fault friction properties by IEAGHG 2015].

Rinehart [2014] found that chlorite-cemented sandstone was weakened on exposure to supercritical $CO_2$. Experimental work of Samuelson and Spiers [2012] examined slip in fault gouges in shale and sandstone and found that the coefficient of friction was not impacted by the presence of $CO_2$, at least in short term experiments. Masoudi et al. [2011] injected shale core with liquid $CO_2$ and found a small decrease in the friction coefficient and small increase in the cohesion, while the Young's modulus increased.

## Experimental Study of Fracture-Permeability in Shale

In this section, we describe experiments that were designed to measure the permeability of fractured shale in order to provide insight into the potential leakage of $CO_2$ from damaged shale caprock. The experiments were conducted with a triaxial coreflood device that simultaneously measures permeability and creates fractures under *in situ* reservoir conditions. The goal was to develop insight into factors (pressure, temperature, stress conditions, shale mineralogy, shale textures) that control the magnitude of fluid leakage through damaged shale.

The experiments were conducted in a custom-designed triaxial coreflood system combined with *in situ* tomography capabilities [Carey et al. 2015; Frash et al. 2016a]. The system is capable of delivering maximum pore pressure and horizontal stress of 34.5 MPa (5000 psi) and maximum axial stress 490 MPa (71,000 psi) and temperature of 100 °C. Fluid injection of water, supercritical $CO_2$, or oil (or a controlled combination of any two phases) is possible. Cylindrical core samples of Utica Shale, courtesy of Chesapeake Energy, were used in the experiments with length and diameter of about 2.5 cm (1") and mineralogy that was roughly 28% carbonate, 22% quartz+feldspar and 50% clay [c.f., Carey et al. 2015]. Experiments were conducted in compression, direct shear and hydraulic fracture configurations (Figure 1).

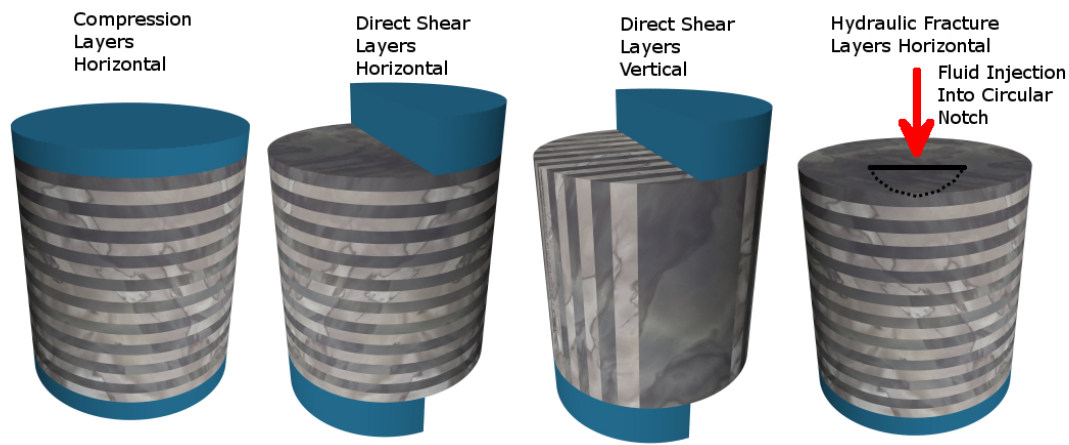

Figure 1. Schematic diagram of compression (left image), direct-shear (two middle images with bedding planes perpendicular on the left and parallel to the load on right), and hydraulic fracture (right image) deformation modes in the triaxial

coreflood experiments. The shale core appears as a striped cylinder with bedding planes either horizontal or vertical. The platens are shown in solid blue.

Compression experiments were based on conventional triaxial methods of squeezing the 1"-diameter specimen between 1"-diameter platens. Direct-shear experiments were conducted with two opposing semi-circular platens that generated a shear plane through the length of the shale specimen. The bedding planes of the shale specimen were oriented either perpendicular or parallel to the axis of the applied load bearing down on the semi-circular platens. Hydraulic fracture experiments were conducted by cutting a circular notch into the face of the shale specimen and injecting at fluid pressures greater than the confining pressure.

The experiments were conducted at 20-50 °C with confining pressures ranging from 3.4 to 22 MPa (500 to 3200 psi). An injection pressure (e.g., 1.4 MPa or 200 psi) using water was maintained at the inlet face of the core sample. Permeability was continuously monitored by a combination of the measured pressure drop and flow rate of water. Deformation was induced by constant flow of the axial pump (resulting in about 10 micro-strains/sec) and measured using a linear variable displacement transducer (LVDT) that recorded displacement of the axial piston. X-ray tomography data were collected as described in Carey et al. [2015] with a resolution of about 25 μm. Measurements of permeability included changing the injection pressure or flow rate as well as the confining pressure, and were conducted over periods ranging from 2 to 5 hours.

**Fracture behavior of shale**

Compression, direct shear, and hydraulic fracture experiments produce distinctly different fracture systems that are strong functions of the orientation of the bedding planes (Figures 2 and 3). Compression experiments produced shear fracture apertures that were generally narrower than those observed in direct shear. Although short sample lengths were used, fractures in compression experiments were not clearly connected between the upper and lower surfaces of the specimens resulting in uncertainty in the permeability measurements. More details on the compression experiments are given in Carey et al. [2014].

Direct shear experiments, in contrast, produced fractures that were well connected between the upper and lower specimen faces and were very suitable for permeability measurements. Fractures in both horizontal and vertical bedding specimens had relatively large apertures resulting in permeability values that were significantly higher than the compression experiments (see below). Hydraulic fracture experiments produced simple fractures that closed upon release of the injection pressure with residual apertures that were less than 25 μm. Additional details on hydraulic fracture experiments are in Frash et al. [2016b].

In direct-shear experiments, fractures that propagated parallel to the bedding planes were simpler (less bifurcations, fewer strands) than fractures that crossed bedding planes (cf., Figures 2 and 3). Carey et al. [2015] interpreted stress-strain behavior to indicate that significant strain in the horizontal-bedding specimen (Figure 3, left) was accommodated by slip along the bedding planes and that this limited the development of through going fractures and permeability.

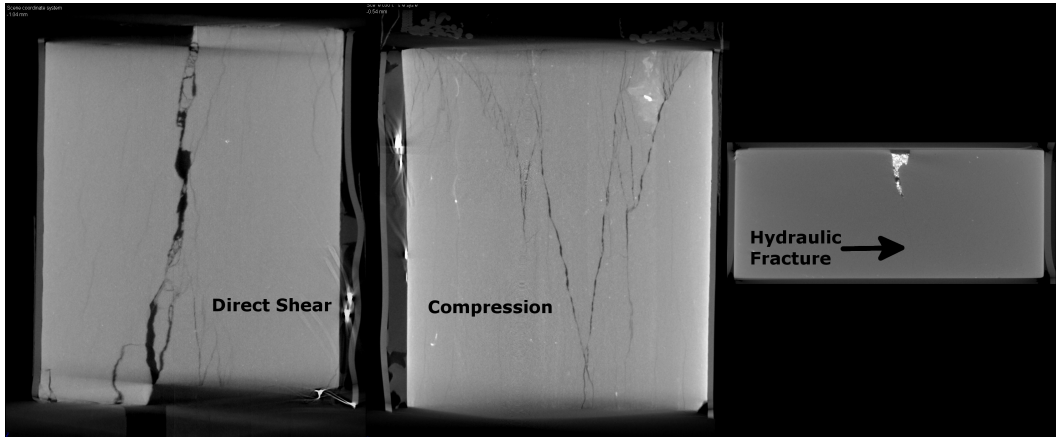

Figure 2. X-ray tomography fracture patterns generated in Utica shale (1"-diameter) using three different triaxial deformation modes. Left: direct shear experiments produce a single dominant, wide-aperture vertical fracture; Middle: compression experiments generate complex, narrow-aperture fractures with top-to-bottom asymmetry; Right: hydraulic fracture experiments produce minute (< 25 µm) single-plane fractures. Bedding planes are vertical in the direct-shear and compression experiments and horizontal in the hydraulic fracture experiment. The hydraulic fracture experiment involved proppants (bright grains at top). Modified from Carey et al. [2015] and Frash et al. [2016b].

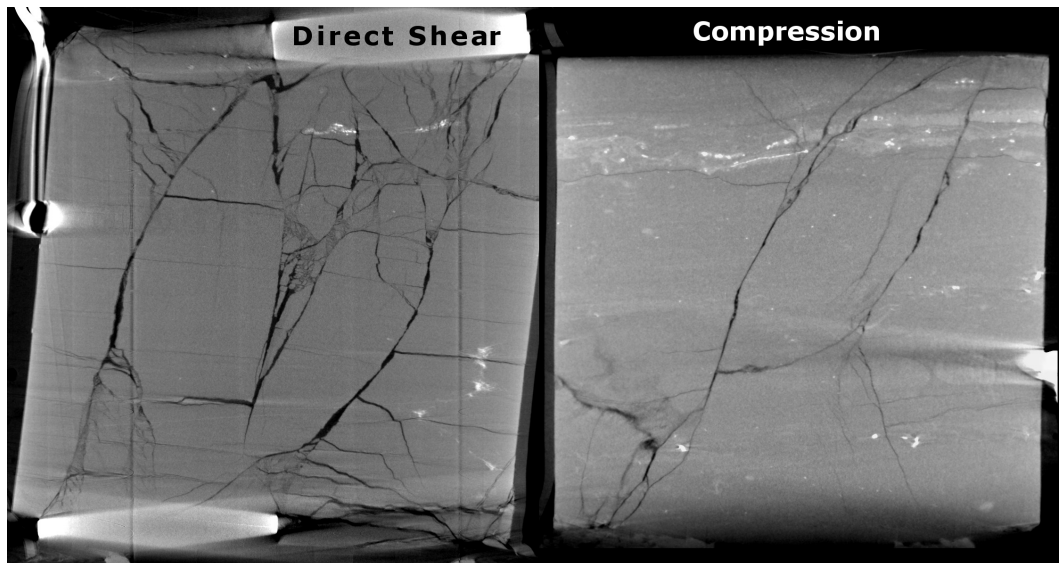

Figure 3. X-ray tomography fracture patterns in Utica shale (1"-diameter) for samples with bedding perpendicular to the load (horizontal with respect to the image) studied in direct-shear (left) and compression experiments (right). Bedding-plane displacements are evident in both tests, while the direct-shear fractures provide good connectivity between the upper and lower faces of the shale specimen. The direct shear platens are visible on the left. Modified from Carey et al. [2014; 2015].

Permeability of fractured shale (referenced to the 1"-diameter core) was continuously measured using water during the fracture experiments as a function of the initial confining pressure (Figure 4). Compression experiments yielded maximum permeability values ranging from 0.1 to 22 mD with fractures parallel to bedding having generally higher permeability. The permeability of direct shear experiments ranged from a low of 30 mD for the horizontal bedding samples to a high of 900 mD for vertical bedding samples [Carey et al. 2015]. The hydraulic fracturing experiment yielded a low permeability value of 0.1 mD [Frash et al. 2016b]. Carey et al. [2015b] also found that permeability and the peak stress achieved before failure (strength) were a strong function of the angle between the vertical bedding plane and the direct-shear plane (Figure 5). Peak strength was a linear function of this angle ranging from 25 to 95 MPa, and the permeability was characterized by a distinct maximum around an angle of 45º between bedding and fracture plane.

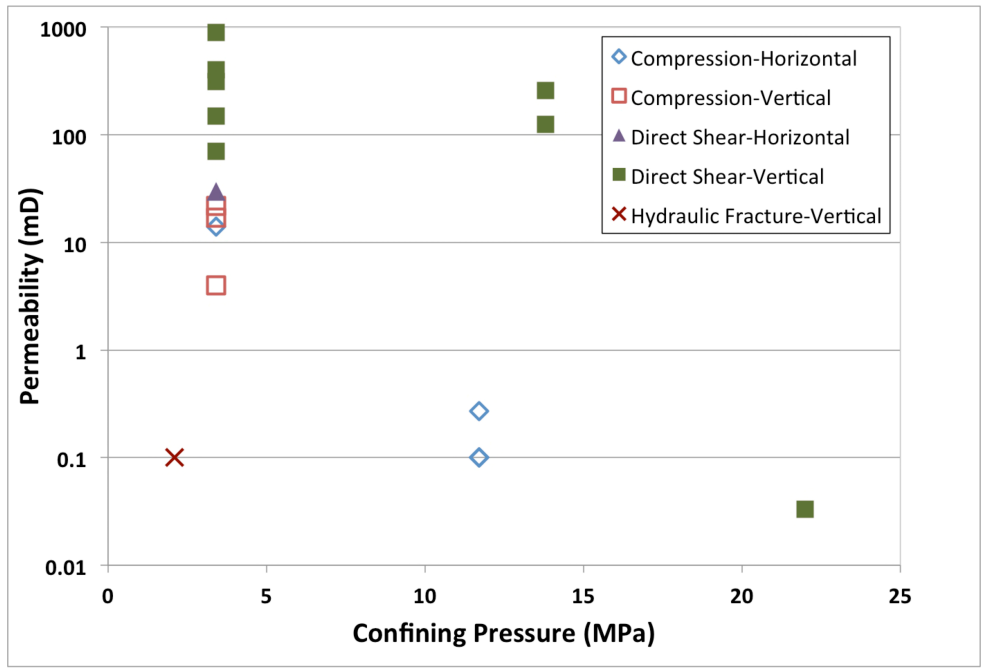

Figure 4. Summary of permeability measurements made on Utica shale using compression, direct shear, and hydraulic fracture methods. Permeability decreases with confining pressure and varies with experiment type.

After formation of fractures, Carey et al. [2015] explored fracture compliance by changing the confining pressure and measuring the resulting changes in permeability. They found an exponential decay in permeability in which $k = P_o\, e^{-aP}$, where $P_o$ was the permeability of the unloaded fracture, P was the confining pressure in MPa, and *a* varied from 0.1 to 0.2. The exponential character of the data suggested that while fracture apertures decreased with applied stress, a return to matrix permeabilities would require quite high stresses.

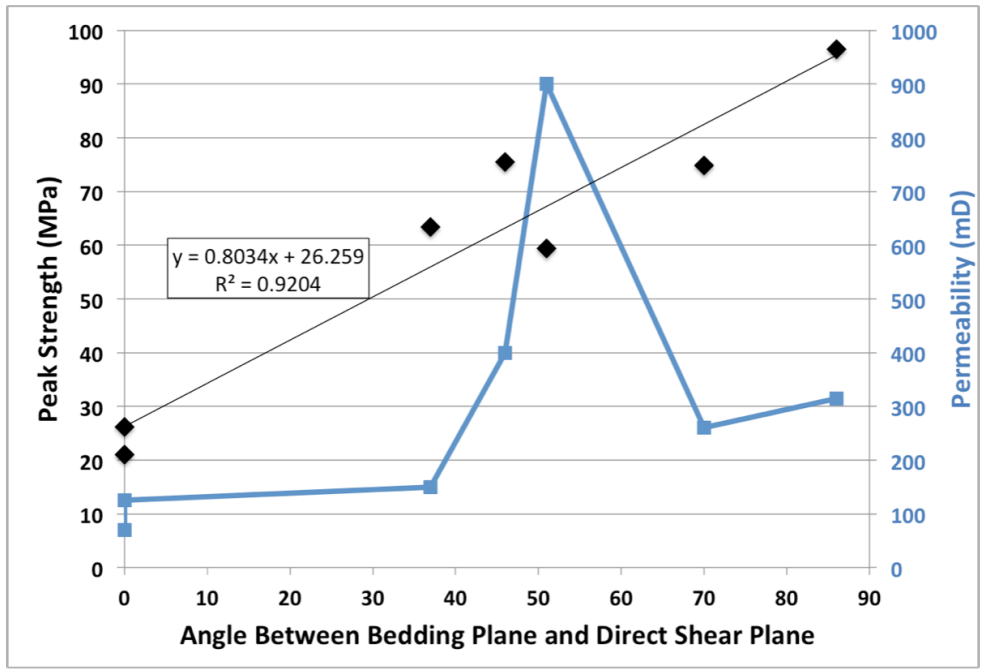

Figure 5. Permeability and strength of Utica shale fractured in direct-shear experiments as a function of the angle between the bedding plane and the shear plane (cf., Figure 1). Modified from Carey et al. [2015b].

In recent work, we have investigated the changing character of fractures observed under *in situ* experimental conditions. The experimental result shown in Figure 4 at 22 MPa confining pressure was particularly interesting. The fractured specimen permeability was only slightly greater than the pre-fracture permeability. *In situ* x-ray tomography showed that despite extensive deformation, fractures were difficult to resolve at the 25-µm resolution of the tomographic images. These results were consistent with a brittle to ductile transition that resulted in dilative, open fracture formation occurring at low pressures (3.4 MPa) and contractive deformation at high pressure (22 MPa). The tomography shows that significant shortening of the sample was accommodated without fracture opening. The permeability measurements support the concept that deformation at these conditions *can* occur without significant development of transmissivity. Upon decompression to atmospheric conditions, a much more significant fracture system developed and had permeabilities that were an order of magnitude greater. The results highlight the challenges in the interpretation of the permeability of fractured materials recovered from high-pressure conditions. Additional details are in Frash et al. [2016a].

# $CO_2$ sorption behavior in shale

Mudrock or shale caprock contains very narrow pores with nanometer dimensions that are the primary cause for their low (matrix) permeability. Most importantly, these nanopores create intimate fluid-rock interactions that could lead to the physical adsorption of gases, i.e., to the trapping of gas molecules at near liquid-like densities attached to surfaces of pores in the material. Thus, $CO_2$ sorption in shale has the potential to mitigate $CO_2$ leakage processes. The mechanism of adsorption is widely exploited for the purification of industrial gas streams, where materials such as activated carbon or zeolites are routinely used that possess large micro- and mesopore volumes and, accordingly, high surface-areas (500-4000 $m^2$/g; Ruthven 1984). In a shale rock, micropores are typically associated with minerals, such as clays, and/or organic material, such as kerogen [Schettler and Parmely 1991]. Micro- (< 2 nm) and meso-pores (2-50 nm) in various shales have been correlated to the dominance of the illite-smectite type of clays in the rock [Kuila and Prasad 2013]; so-called intra-particle organic pores with diameters as small as 4 nm have been observed in samples from the Barnett Shale and contribute to grain porosities as high as 20-30% [Loucks et al. 2009]. Remarkably, some results suggest that 50-80% of the total amount of natural gas found in shale plays is trapped as an adsorbed phase in the micropores of the rock [Curtis 2002]; however, other studies suggest that the contribution of adsorbed gas to total production may be small when the latter is driven by a reduction in reservoir pressure only [Heller and Zoback 2014].

To better appreciate the extent of gas sorption on shale, we compare $CO_2$ adsorption isotherms measured at 50°C and pressures up to 20 MPa on two commercially available sorbent materials (activated carbon and zeolites), a bituminous coal, and a shale (Figure 6). The adsorption capacity of "engineered" materials is significantly larger (~30-37 wt%) than the values obtained for natural materials (2 and 8 wt% for shale and coal, respectively). However, as shown by the filled symbols in the same figure, this margin is greatly reduced when the total volume of the system (e.g., the fixed-bed adsorption column or the geologic unit) is considered, thus highlighting the significance of adsorption in (dense) microporous rocks. Accordingly, we argue that to fully evaluate the sealing effectiveness of (damaged) shale caprock, a better understanding is needed of the mechanisms of adsorption.

Measuring adsorption at conditions representative of shale caprock requires achieving elevated pressures (> 10 MPa) and temperatures (> 40°C). While under these conditions the behavior of the free gas phase can be described by an equation of state, the thermodynamic treatment of the adsorbed phase is more complex. For a gas below the critical point (for $CO_2$ this is 31.1 °C, 7.39 MPa), the adsorption mechanism is controlled by the vapor-liquid equilibrium and, accordingly, the density of adsorbed fluid takes a value near that of the saturated liquid. On the other hand, supercritical fluids show no discontinuities in the transition to a dense phase, and the density in its adsorbed state is not well defined. The debate regarding the mechanism of adsorption and its appropriate representation under such conditions are still matter of controversy [Zhou and Zhou 2009]. A recent publication that reviews various adsorbate/adsorbent systems shows that values for the adsorbed density can range between the liquid and the solid state of the adsorbate even for similar pressure and temperature conditions, and large discrepancies were found between different measurement techniques [Pini 2014a]. In practice, because the molecular volume of the adsorbed phase cannot be determined accurately in microporous material, *incremental* rather than *absolute* quantities are used to report adsorption at elevated pressures [Gumma and Talu 2010]. Such measures require the definition of a reference state; when the skeletal (inaccessible to adsorption) volume of the solid material is known, the so-called excess adsorbed amount is obtained from an adsorption measurement that is defined as the difference between the actual amount adsorbed ($n^A$) and the amount of homogeneous bulk fluid with density $\rho_M$ that would be present in the (unknown) volume occupied by the adsorbed phase ($V^A$; i.e. as $n^{EX} = n^A - \rho_M V^A$; Sircar 2001). We refer here to the skeletal volume as the impenetrable volume of the adsorbent, i.e. the value obtained from a pycnometer measurement with helium. Yet, concerns are being raised with regards to finding an appropriate methodology to measure the inaccessible volume of a microporous material [Do and Do 2007; Do et al. 2010; Gumma and Talu 2010; Pini 2014a], as this may significantly affect the behaviour of the adsorption isotherm at elevated pressures (or densities), a condition that is indeed readily achieved in the subsurface.

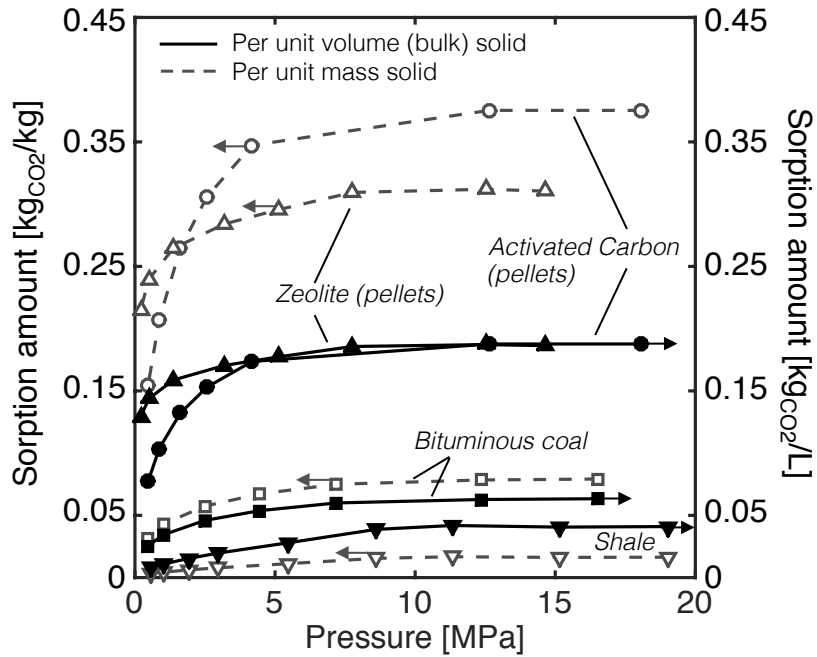

Figure 6. $CO_2$ sorption on various sorbent materials at 50°C as a function of gas pressure: activated carbon [Pini et al. 2006; Filtrasorb 400, Chemviron Carbon, $\rho_B$=0.5g/cc; $V^A$=0.344 cc/g], 13X zeolite [Pini 2014a; Z10-03, Zeochem, $\rho_B$ = 0.6 g/cc; $V^A$ = 0.27 cc/g)], a bituminous coal [Pini et al. 2010; Australia - A1, $\rho_B$ = 0.8g/cc; $V^A$ = 0.055 cc/g] and a shale (this study, $\rho_B$ = 2.5 g/cc).

As an example of general validity, $CO_2$ adsorption data on a dry Eagle Ford Shale sample are presented in Figure 7 that have been measured at 50°C and over a wide range of pressure (0-20 MPa) in a Rubotherm magnetic suspension balance. The grey-filled circles are raw excess adsorption data obtained by using a solid volume, $V^S$, of the shale measured by a (traditional) Helium experiment ($\rho^S$ = 2.87 g/cm$^3$); the empty symbols refer to the same data, but where $V^S$ has been obtained by XRD (circles, $\rho^S$ = 2.64 g/cm$^3$, by assuming values for the density of the main mineral constituents (calcite-quartz-clays) obtained by quantitative x-ray diffraction and a kerogen density of 1.3 g/cm$^3$) or by choosing its value such that the isotherm is positive (squares, "non-zero excess") over the entire pressure range ($\rho^S$ = 2.59 g/cm$^3$). As expected from the definition of excess adsorption, the isotherms are characterized by a maximum and then decrease linearly when sufficiently large fluid densities are reached, i.e. when adsorption saturation is likely to be attained. It can be seen that the choice of the parameter $V^S$ affects significantly the shape of the isotherm, a result that has been reported also for activated carbon [Malbrunot et al. 1997] and zeolite [Pini 2014a]. In the case of a weakly adsorbing material, such as shale, this effect is exacerbated and leads to a significant portion of the adsorption isotherm having negative values. While the existence of negative adsorption values has been questioned in the literature [Malbrunot et al. 1997; Do et al. 2010], the definition of excess adsorption as an incremental quantity doesn't necessarily preclude it [Gumma and Talu 2010]. In fact, one could argue that at large densities, steric effects associated with the confined space of a micropore lead to a less effective packing of the molecules as compared to the (free) bulk phase. Nevertheless, the discrepancy between the three sets of data raises concerns with respect to the interpretation of adsorption experiments at elevated pressures with materials that possess relatively low adsorption capacity. Absolute sorption values, $n^A$, can be calculated by estimating the volume of adsorbed phase, $V^A$, from the slope of the linear region of the isotherm, i.e. when adsorption saturation has been reached, and by assuming that this value is independent of pressure. For the three excess isotherms reported, $V^A$ takes a value of 0.06, 0.03 and 0.02 cm$^3$CO$_2$/g rock, respectively. Note that upon forcing the adsorbed phase to be constant in volume, we assume that it entirely fills the (micro)pore space, irrespectively of the gas pressure. As shown by the black-filled symbols in the figure, this graphical approach is not affected by the choice of the parameter $V^S$, thus increasing the reliability of these estimates. For the Eagle Ford sample considered here, $CO_2$ sorption increases up to a saturation value of about 0.4 molCO$_2$/kg (~300 SCF/ton); in comparison, $CO_2$ adsorption capacities in the range 0.7–0.8 molCO$_2$/kg and 0.7–2 molCO$_2$/kg have been observed at similar conditions for samples of Devonian and Permian shale, respectively [Weniger et al. 2010].

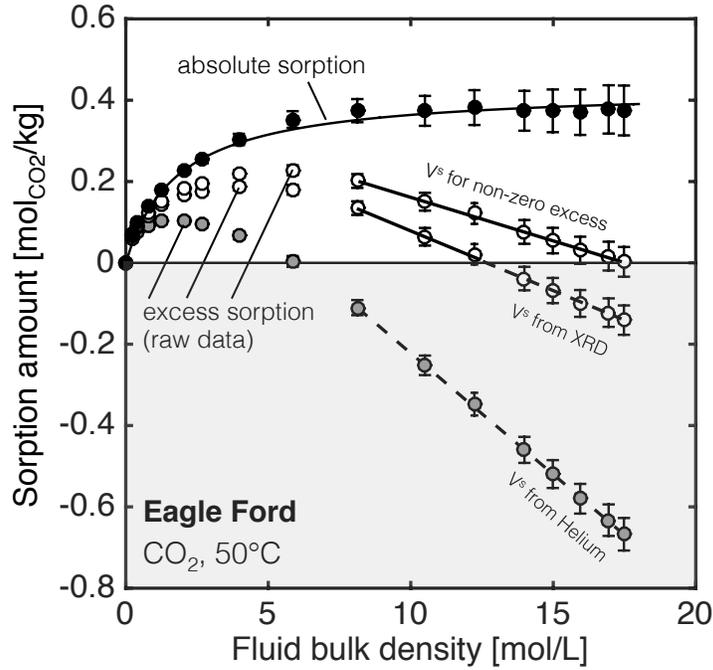

Figure 7. Experimental data set from a high-pressure $CO_2$ adsorption experiment carried out at 50°C on an Eagle Ford sample (dried at 105°C, size fraction: 180-425 μm). The sample has a TOC content of about 7 wt% and a total pore volume of 0.04 $cm^3$/g (as measured by $N_2$ adsorption at 77K). Three sets of excess adsorption isotherms are shown that differ in the value assumed for the skeleton volume of the sample; absolute sorption data (black filled symbols) are estimated by following a graphical approach that uses the slope of the linearly descending part of the isotherm.

Mathematically, the fluid storage capacity of a caprock can be expressed as:

$$S_{CO_2} = \rho_{CO_2} \left[ \frac{\phi}{\rho_B} - V^A \right] + n^A \qquad (1)$$

where $S_{CO_2}$ (mol-$CO_2$/kg-rock) is the gas-in-place per unit weight of material, $\rho_{CO_2}$ (mol-$CO_2$/L-$CO_2$) is the gas-phase molar density, $\rho_B$ (kg rock/L bulk rock) is bulk rock density, $V^A$ (mol-$CO_2$/L-rock) the adsorbed-phase volume, $\phi$ (-) is the total (effective) porosity, and $n^A$ (mol-$CO_2$/kg-rock) is the adsorbed amount. The first term on the right-hand side of Eq. 1 represents the amount of free gas in the pore space of the rock (described by an equation of state, such as a real gas law) and accounts for the reduction of the pore space due to gas adsorption. The second term is the amount of adsorbed gas, which can be obtained through the measurement of an adsorption isotherm, such as the one presented in Figure 7. Application of Eq. 1 to the Eagle Ford sample considered here leads to the situation depicted in Figure 8, where the total $CO_2$ uptake (converted to SCF/ton) is plotted as a function of the fluid pressure. In the figure, the dashed and dash-dot curves represent the total storage capacity of the rock, as given by either one of the two mechanisms alone. As expected, adsorption is particularly effective below 1200 psia (pounds per square inch absolute; as the adsorbed phase is still much denser than the gas phase), while the contribution of free gas increases with pressure and eventually overcomes adsorption. Unfortunately, common practices to estimate volume capacities often improperly account for gas adsorption by neglecting the volume occupied by the adsorbed fluid (i.e., by assuming $V^A = 0$ or $\rho^A \to \infty$; Ambrose et al. 2012). As shown in the figure, such an approach, which is obtained as the simple sum of gas compression and gas sorption (dashed and dash-dot lines in Figure 8, respectively), leads to a significant overestimation of the storage capacity. Yet, when $V^A$ is accounted for, assumptions are made regarding its value (or equivalently the adsorbed density). As discussed by Pini [2014b], values between 600 and 1400 kg/m$^3$ have been reported in the literature for the adsorbed gas density observed in clay and shale samples. We have shown in Figure 7 that the method of estimating the adsorbed phase volume from the measured high-pressure adsorption isotherms creates an additional source of uncertainty ($\rho^A$ = 560 - 700 kg/m$^3$). The solid curves in Figure 8 represent total storage capacity ('total uptake') estimates based on four distinct values of $\rho^A$ (or $V^A$), namely $\rho^A$ = 560, 700, 1200 kg/m$^3$ and $\rho^A \to \infty$, with the area shaded in purple referring to the region where most of the estimates are found. It can be seen that knowledge of the density of the adsorbed phase is quite important as its effect on the total uptake is quite large; also, in agreement with results reported above, the contribution of adsorption is dominant below about 1500 psi and the appearance of a negative excess adsorbed amount (the case with $\rho^A$ = 560 kg/m$^3$) leads to a total uptake capacity at elevated pressures that is less than the amount that would be stored by simple gas compression without adsorption.

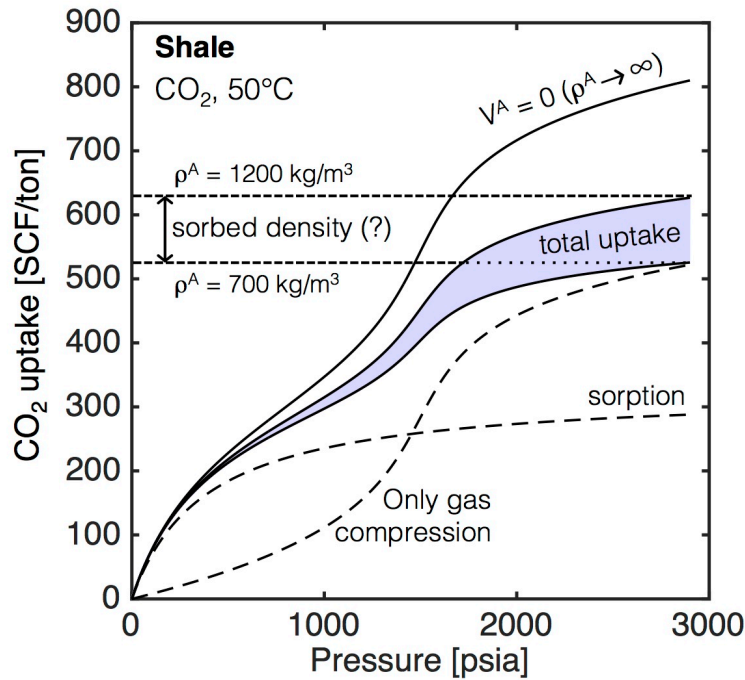

Figure 8. Theoretical $CO_2$ storage capacity of a shale caprock based on experimental data from a sample of Eagle Ford shale. The dashed and dash-dot lines represent contribution from gas compression in the entire pore space ($\phi \approx 9\%$) and from gas adsorption in the micropores (described by a curve fitted to the black-filled symbols in Figure 7), respectively. The solid curves have been estimated from Eq. 1 by assuming distinct values of the adsorbed phase density (or, accordingly, volume), namely $\rho^A$ = 560, 700, 1200, $\infty$ kg/m$^3$.

## Acoustic properties of the $CO_2$-shale system

One approach to monitoring and characterizing $CO_2$ infiltration into caprock utilizes changes in the acoustic properties of $CO_2$-saturated rock. Numerous studies have been conducted to study the effect of replacing water or hydrocarbon with $CO_2$ in potential reservoir rocks [for example, Batzle et al. 2005; Adam et al., 2006; Vanorio et al. 2010; Yam and Schmitt 2011; Sharma et al., 2013;Njiekak et al., 2013]. Changes in seismic P- and S-wave velocities due to rock-fluid interactions have been used to develop poroelastic models of rocks [Dvorkin and Nur, 1993; Berryman and Wang, 1995; Diallo et al., 2003; Khazanehdari and Sothcott 2003]. These studies have been able to detect changes in seismic signatures due to displaced reservoir fluids in mostly clastic reservoirs. Application of these rock physics methods to caprocks has been more limited, largely due to the limited experimental data on mudrocks. As compared with conventional reservoir rocks, most caprocks are characterized by a relatively higher degree of heterogeneity of the pore topology as well as mineral composition, containing a mixture of clay, quartz, and carbonate minerals and organic matter. This may result in a non-uniform rock-fluid interaction and heterogeneous saturation profile in the pores due to the preferential coverage of different fluids for specific rock components [Sharma et al., 2013]. For example, Kumar et al. [submitted] have shown that organic-rich shales are characterized by selective adsorption of water in the inorganic clay pores and not the organic pores. Thus, in a multi-mineralic rock, water is likely to selectively saturate clay bound pores, while the organic pores will be saturated by the non-aqueous phase.

Here, we consider the isotropic formulation of the theoretical Gassmann fluid substitution to quantify the effect of preferential fluid saturation in organic-rich caprock. Kumar et al [submitted] have quantified mesopore and micropore volumes of various storage and caprocks for different sorbents (Table 1). We hypothesize a case in which $CO_2$ is injected in a water-saturated reservoir overlain by a water-saturated caprock. With $CO_2$ invasion, the new saturation profile in the caprock would be as follows. At the injection pressure, $CO_2$ can be expected to saturate the pores of the caprock that are comparable in size with the pores in the reservoir. However, pore sizes are much smaller in caprocks [Kuila and Prasad, 2013] and are dominated by meso- and micro-pores. Thus, the displacement of the aqueous phase from the smaller pores with the invading $CO_2$ will take place only if $CO_2$ pressure is larger than the capillary pressure of the $CO_2$-aqueous phase in the pores. The capillary pressure, in turn is dependent upon the degree of affinity of the pores towards the aqueous phase. This preference was determined by quantifying separate water and hexane vapor adsorption experiments; the results obtained are listed in Table 1. The porosity obtained from water vapor is smaller than that with hexane vapor, which reflects the saturation preference of organic matter for the non-aqueous fluid. Similar to the preferential hexane saturation, we postulate that the initially water-saturated meso- and macro-pores should be displaced by the injected $CO_2$. Thus, in our hypothesis, the patchy saturation profile after $CO_2$ injection can be summarized as follows:

1) Macro-sized pores occupied by gaseous $CO_2$ phase
2) Clay bound meso and micro-pores occupied by water. $CO_2$ can only exist in these pores if it is dissolved in the water.
3) The organic matter pores (difference between the hexane-derived and water-derived pores) are occupied with $CO_2$. This $CO_2$ can be supercritical if the conditions exceed the critical point of $CO_2$. As a result, the organic meso-pores will be filled with supercritical $CO_2$ gas, whereas the organic micropores are filled with an adsorbed $CO_2$ phase with a density of 1.03g/cm³ corresponding to that of the val der Waals volume [Humayun and Tomasko, 2000], which is 23.45 mmol/cm³.

In our fluid substitution model for predicting the seismic properties of caprock, we followed the above mentioned saturation profile. We ignored the effect of $CO_2$ dissolved in the aqueous phase because of the low solubility of $CO_2$ in water giving negligible changes in the acoustic properties of water. Figure 9 shows the significant reduction in bulk modulus of the caprock after $CO_2$ invasion.

**Table 1**. Effect of $CO_2$ invasion on the bulk modulus (K) of caprock modeled using Gassmann fluid substitution.[a]

| Sample | $\emptyset_{Hi+Ho}$ | | | $\emptyset_{Hi}$ | | | $\emptyset_{Ho}$ | | | $K_o$ | $K^*$ | $K_{init}$ | $K_f^{ads}$ | $K_f^{gas}$ | $K_{final}^{ads}$ | $K_{final}^{gas}$ |
|---|---|---|---|---|---|---|---|---|---|---|---|---|---|---|---|---|
| | Mes | Mic | Tot | Mes | Mic | Tot | Mes | Mic | Tot | | | | | | | |
| UBS1 | 8.6 | 1.8 | 10.4 | 5.5 | 0.8 | 6.3 | 3.0 | 1.1 | 4.1 | 32.5 | 27.7 | 35.7 | 0.46 | 0.42 | 29.3 | 29.1 |
| UBS5 | 6.1 | 1.3 | 7.4 | 3.3 | 0.4 | 3.7 | 2.8 | 0.9 | 3.8 | 34.4 | 30.8 | 43.4 | 0.38 | 0.35 | 32.6 | 32.4 |
| UBS13 | 7.9 | 1.5 | 9.4 | 1.4 | 0.2 | 1.7 | 6.4 | 1.3 | 7.8 | 37.3 | 32.2 | 42.3 | 0.24 | 0.23 | 33.2 | 33.1 |
| LBSG | 9.0 | 2.2 | 11.2 | 3.6 | 0.3 | 3.9 | 5.4 | 1.9 | 7.4 | 37.1 | 31.1 | 39.2 | 0.30 | 0.28 | 32.6 | 32.0 |
| 7219 | 13.5 | 2.5 | 16.0 | 8.9 | 0.5 | 9.4 | 4.5 | 2.0 | 6.6 | 30.7 | 23.7 | 28.6 | 0.46 | 0.41 | 24.7 | 24.6 |
| 11005 | 8.0 | 3.6 | 11.6 | 2.4 | 0.4 | 2.8 | 5.5 | 3.2 | 8.8 | 34.3 | 28.6 | 36.0 | 0.28 | 0.24 | 29.5 | 29.4 |

[a]Estimates of hydrophilic ($\emptyset_{Hi}$) and hydrophobic ($\emptyset_{Ho}$) meso- and micro-pores were obtained from hexane and water vapor adsorption. The initial bulk modulus for water saturated rock ($K_{init}$) and the final bulk modulus after invasion with $CO_2$ ($K_{final}$) were calculated by using a frame bulk modulus ($K^*$) that was a weighted average of the mineral bulk moduli ($K_o$); a Reuss average was used for the fluid bulk modulus ($K_f$) (Smith et al., 2003) after $CO_2$ invasion. The effect of supercritical $CO_2$ sorption in micropores was modeled assuming that the $CO_2$ fluid modulus in hydrophobic micropores was the same as in the condensed liquid phase at near

critical conditions (0.29 MPa). In the remaining pores, it was assumed to be the same as that of its gas phase (0.19 MPa).

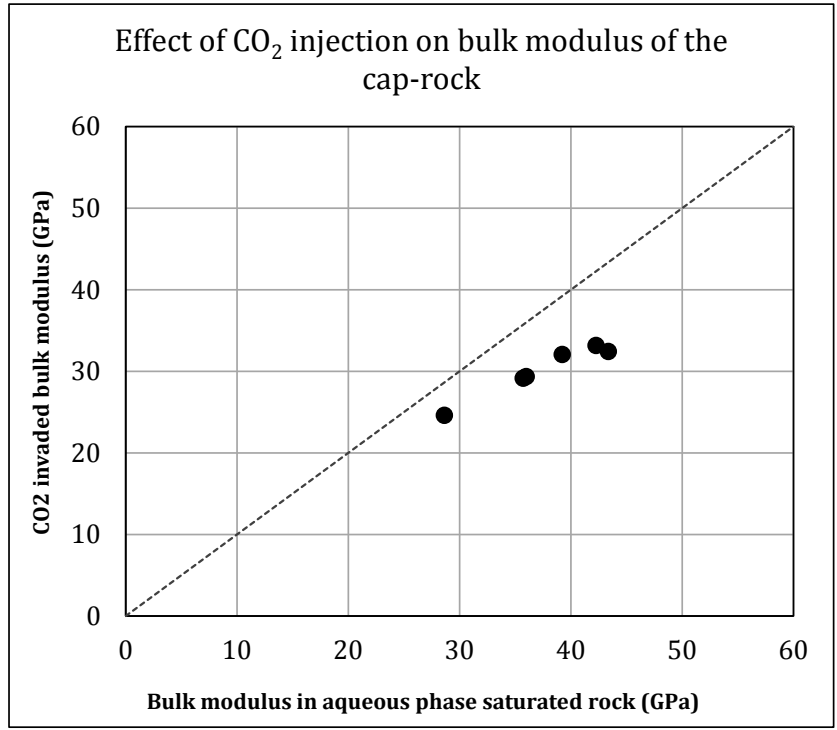

Figure 9. Comparison of the bulk moduli of initial and final bulk modulus of the cap-rock. The bulk modulus can be reduced significantly with $CO_2$ invasion in the cap rock.

In the above analysis, $CO_2$ is considered as an inert gas that does not interact with the mineral matrix. However, core damage after $CO_2$ injection has been quantified in numerous rocks [for example, Browne, 1978; Adam et al., 2006; 2013; Andreani et al., 2009; Sharma, 2015]. In carbonate rocks, core damage due to dissolution is followed by carbonate reprecipitation [Sharma, 2015]. Formation damage due to $CO_2$ interaction with the caprock has largely been ignored. Given the multi-mineralic nature of most caprocks, preferential surface coverage depending on mineral-fluid interactions could lead to selective mineral alterations depending on $CO_2$ access and reactivity with the host minerals. Thus, caprock might experience selective damage depending on fluid access and coverage that could result in additional changes to shale caprock acoustic properties.

## Conclusions

Caprock integrity is one of the chief concerns in the successful development of a $CO_2$ storage site. In this chapter, we discuss fracture-permeability relationships in shale, $CO_2$ sorption by shale, and the acoustic properties of $CO_2$-saturated shale. These sections provide an overview of the permeability of fractured shale, the potential for mitigation of $CO_2$ leakage by sorption to shale, and the detection by acoustic methods of $CO_2$ infiltration into shale.  A review of the literature reveals that while significant concerns have been raised about the potential for induced seismicity to damage the caprock and cause leakage, relatively little is known about the permeability of the damaged shale. Previous studies, motivated in part by the role of shale as caprock to hydrocarbon reservoirs, have identified a critical transition between brittle and ductile deformation of shale that corresponds to a change from dilation to contraction during shear deformation. This transition is clearly a function of temperature, pressure, previous burial/diagenetic history, and the shale mineralogy and textural features.

Shale that deforms in the ductile regime would be expected to have significantly lower permeability than shale experiencing brittle fracturing. Knowledge of the permeability of damaged shale in both regimes is necessary if we are to evaluate the potential consequences of induced seismicity. Relatively little is known of this permeability either in the field or laboratory. We present a summary of recent experimental work conducted by the authors that show profound differences in permeability of up to three orders of magnitude between brittle and ductile fracture permeability. In the ductile regime, it is possible that shale caprock could accommodate deformation without a significant loss of $CO_2$ from the storage reservoir.

Our work also examined the effect of changes in confining pressure on the permeability of fractured shale. The results show a modest degree of permeability reduction with increased pressure and provide a measure of fracture compliance. Perhaps of even greater interests is the question of creep and self-healing of fractured shale. Studies conducted as part of the European nuclear waste disposal program clearly show strong evidence of self-healing in favorable shale types. In the absence of long-term experiments on creep, the experiments on the effect of confining pressure may serve as a proxy for the potential of creep to reduce permeability. Presumably, fractures with greater compliance would also be likely to experience greater creep over long periods of time.

In cases where $CO_2$ does migrate through damaged shale caprock, $CO_2$ sorption onto shale mineralogy may have a mitigating impact. Measured adsorption values on shale reported here range from 0.4 to 2 mol-$CO_2$/kg-shale_ and are significantly smaller than $CO_2$ sorption onto commercial sorbents. However, the tremendous mass of shale available in the subsurface implies that significant $CO_2$ may be captured by shale. The total storage capacity of caprock is the sum of the free gas phase present in the pores and the adsorbed gas on the pore walls. Adsorption is dominant at lower pressures (below about 10 MPa) and the free phase becomes more important as pressure and the density of the free phase increase. Measured total storage capacities range from 1 to 45 kg-$CO_2$/tonne-shale.

Once $CO_2$ is in the caprock, changes in the acoustic properties of $CO_2$-saturated shale might provide a means of detecting or monitoring $CO_2$ leakage. Although this approach has been applied extensively to reservoir rocks (porous sandstones and carbonates), application to caprock with elastic anisotropy is much more limited.

Theoretical work described here suggests that during $CO_2$ invasion of caprock, macro-sized pores and organic-lined meso- and micro-pores would be filled with $CO_2$, while meso- and micro-pores lined by clay would retain water. The resulting Gassmann fluid substitution calculations show a significant reduction of the bulk modulus of $CO_2$-saturated shale. However, the impact of reaction-induced changes to the acoustic properties of shale has yet to be determined.

The recent development of shale gas resources has created an opportunity to focus greater attention on the fracture and fluid-flow properties of shale and how shale performs both as a hydrocarbon resource and as caprock. In our summary of the literature and presentation of recent experimental results, it is clear that much remains to be learned of the permeability of fractured shale, the conditions under which $CO_2$ sorption in shale mitigates $CO_2$ leakage, and the detection of $CO_2$ leakage through shale. There are a wide variety of shale types and developing a predictive capability in all three areas is a grand challenge. One of the key questions is fracture compliance and the time-scale for creep and self-healing to occur and shutdown the permeability. To the extent that this time scale is short, $CO_2$ leakage from damaged caprock could be temporally limited. In addition, other lines of evidence indicate that fluid flow through caprock may shutoff below a critical fluid pressure necessary to maintain fracture apertures and permeability.

## Acknowledgements

This work is supported by the U.S. Department of Energy (DOE) National Energy Technology Laboratory (NETL) under Grant Number DEFE0023223, which is managed and administered by the Colorado School of Mines and funded by DOE/NETL and cost-sharing partners, and by Grant Number FE-715-16-FY17, which is managed and administered by Los Alamos National Laboratory. We sincerely thank the U.S. Department of Energy for funding this research.